\documentclass[twocolumn,floatfix,amsmath,amssymb,prb,showpacs,superscriptaddress]{revtex4-1}
\usepackage{graphicx}

\usepackage{dcolumn}
\usepackage{bm}
\usepackage{amsmath}
\begin{document}
\title{Strain-induced modifications of transport in gated graphene nanoribbons}
\author{Diana A. Cosma}
\email[Electronic mail: ]{d.cosma@lancaster.ac.uk}
\affiliation{Department of Physics, Lancaster University, LA1 4YB Lancaster, United Kingdom}
\author{Marcin Mucha-Kruczy\'{n}ski}
\affiliation{Department of Physics, University of Bath, Claverton Down, Bath, BA2 7AY, United Kingdom}
\author{Henning  Schomerus}
\affiliation{Department of Physics, Lancaster University, LA1 4YB Lancaster, United Kingdom}
\author{Vladimir I. Fal'ko}
\affiliation{Department of Physics, Lancaster University, LA1 4YB Lancaster, United Kingdom}
\date{\today}
\begin{abstract}
We  investigate the effects of homogeneous and inhomogeneous deformations and edge disorder on the conductance of gated graphene nanoribbons. Under increasing homogeneous strain the conductance of such devices initially decreases before it acquires a resonance structure, and finally becomes completely suppressed at larger strain. Edge disorder induces mode mixing in the contact regions, which can restore the conductance to its ballistic value.
The valley-antisymmetric pseudo-magnetic field  induced
by inhomogeneous deformations  leads to the formation of additional resonance states, which either originate from the coupling into Fabry-P\'{e}rot states that extend through the system, or from the formation of states that are localized near the contacts, where the pseudo-magnetic field is largest. In particular, the $n=0$ pseudo-Landau level  manifests itself via two groups of conductance resonances close to the charge neutrality point.
\end{abstract}

\maketitle

\section{Introduction}
Monolayer graphene~\cite{castro-neto1} is a unique material capable of sustaining reversible deformations in excess of several percent.~\cite{castro_neto,ribeiro,liu_ming,kim1,strong} The effects of strain in this one-atom-thick crystalline membrane~\cite{membrane1,membrane2} attract attention due to the peculiar way in which they affect the already unusual electronic properties of this material. \cite{castro-neto1,effects,effects2}
Pristine graphene displays a conical dispersion~(Dirac points, DPs) at the gapless edge between the valence and conduction bands.
The DPs are replicated at the inequivalent $K$ and $K'$ corners of the hexagonal Brillouin zone (BZ), and the effect of lattice deformations on electrons is equivalent to that of an effective gauge field with the sign inverted in the opposite valleys. \cite{koshelev,castro_neto,ribeiro,guinea,vozmediano,rainis} Consequently, homogeneous deformations result in a small shift of the Dirac cones from the corners of the BZ, \cite{castro_neto} whereas inhomogeneous strain influences electron motion similarly to a valley-dependent effective pseudo-magnetic field. \cite{suzura,prada,manes2,guinea,low,effects,pereira} Recent scanning-tunneling experiments on graphene nanobubbles~\cite{crommie} revealed that even small inhomogeneous deformations can induce pseudo-magnetic fields that reach values equivalent to hundreds of Teslas. Such strong fields result in the localization of the electronic states, and lead to the formation of a discrete `Landau level'~(LL) spectrum with the peculiar $n=0$ LL state positioned at zero Fermi energy~($E_\mathrm{F}=0$). \cite{guinea,suzura,prada,manes2,crommie,schomerus,poli}

In this paper, we perform a systematic analysis of the conductance of gated armchair graphene nanoribbons~(GNRs), which are  subjected to both homogeneous and inhomogeneous longitudinal deformations, as well as to various types of edge disorder.
Our calculations are carried out within a tight-binding model that incorporates the strained-induced modifications of the couplings. \cite{marcin}
The conductance is then obtained in the Landauer-B{\"u}ttiker approach, \cite{buttiker}
where the transmission probabilities are obtained using the recursive Green's function technique. \cite{datta,henning1}

Under homogeneous deformations of increasing strength, the conductance of such ribbons first decreases, then acquires a resonant structure, and finally becomes completely suppressed in a large range of energies. These effects arise from a combination of a strain-induced  mismatch of the Fermi surfaces in the leads and the strained regions, \cite{pereira} and the finite-size quantization of the transverse momentum.
We found that these transport features are washed out by single-atom edge defects, while double-atom defects (consisting of the removal of a dimer at the edge) do not alter the resonant structure significantly, and even can restore the ballistic transport properties of the ribbon in the regime where the conductance is completely suppressed by the deformations in the absence of disorder.

Suspended graphene nanoribbons\cite{suspended,suspended2,with_strain,with_strain2,with_strain3} display inhomogeneous strain distributions. \cite{suzura,prada,manes2} We show that in this case the pseudo-magnetic field gives rise to a characteristic set of additional resonances.
The nature of these resonances is revealed through the local density of states~(LDOS) profiles, which we calculate at the resonance energies. We found that these features can be attributed either to Fabry-P\'{e}rot-like standing waves, or to resonant transmission via pseudo-magnetic Landau level states that form in the contact regions of the GNR. \cite{gradinar}
The 0th LL is identified by its sublattice polarization, \cite{guinea,poli} and is found to result in resonance states close to the charge neutrality point.

The abovementioned results are described in detail in Sec.~\ref{sec:conductance}. The preceding Sec.~\ref{sec:hamiltonian} introduces the tight-binding model for strained graphene ribbons and identifies the underlying physics of homogeneously and inhomogeneously strained armchair GNRs, while Sec.~\ref{sec:conclusions} summarizes their consequences.

\begin{figure}[t!]
\includegraphics[width=\columnwidth]{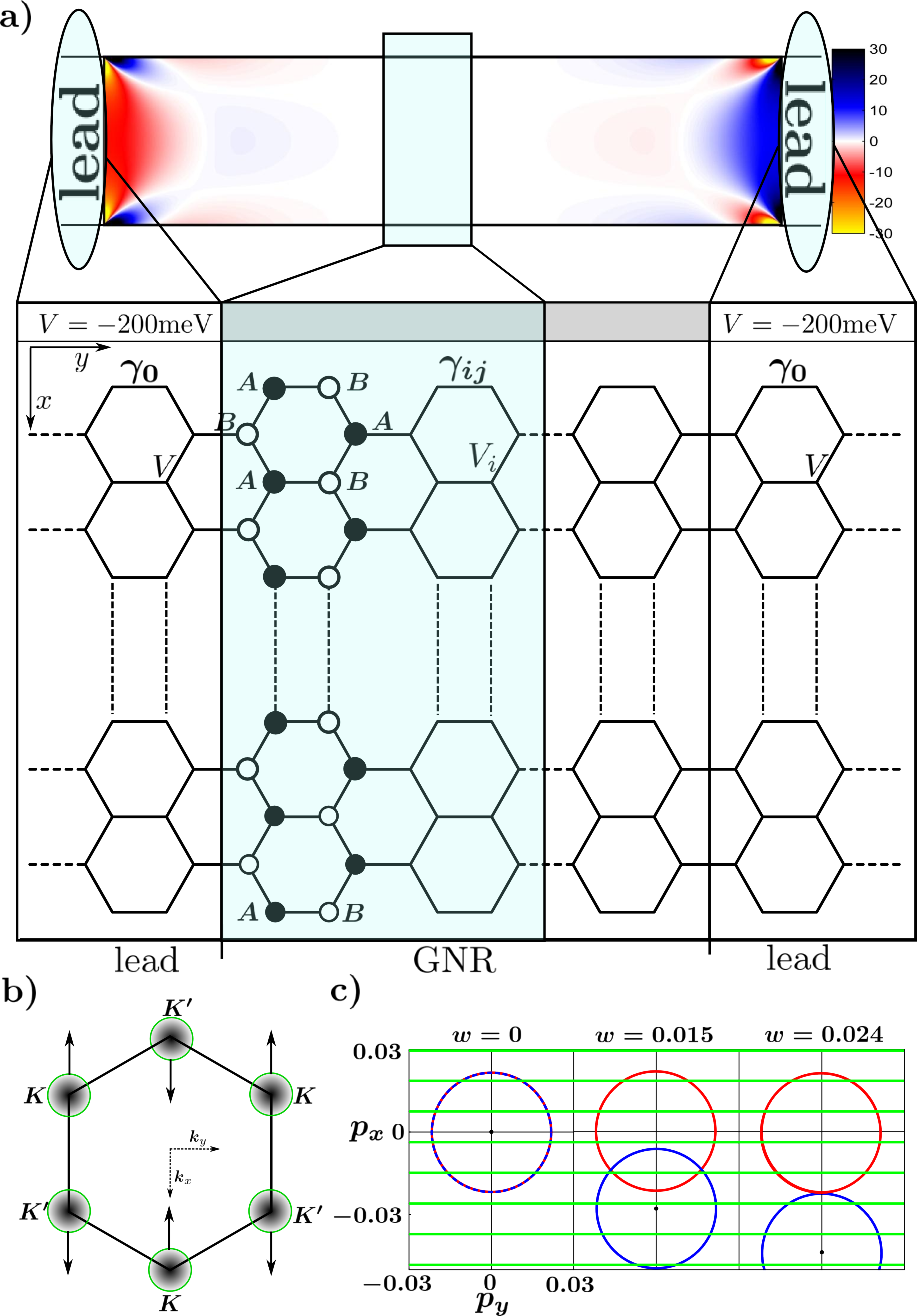}
\caption{\label{fig1}(Color online) Panel (a) shows a sketch of a graphene nanoribbon (GNR) of  aspect ratio $L/W=4$, where $L$ is the length and $W$ is the width of the system. The color code indicates the pseudo-magnetic field $\mathcal{B}(\mathrm{Tesla})$ for electrons in the $K$ valley, at $w=0.05$ inhomogeneous tensile strain in the middle of a system of width $W\simeq40$nm.
We also sketch the honeycomb lattice corresponding to the tight-binding model in Eq.~(\ref{eqn-hamiltonian}). The leads are heavily doped by imposing an onsite potential $V=-200$meV. This potential step can be controlled via electrostatic gates. In the central region the strain modulates the hopping matrix elements $\gamma_{ij}$ and the on-site energy $V_i$. b) Shift of the Dirac cones from the corners $K$ and $K'$ in the Brillouin zone of a homogeneously strained armchair GNR. c) Fermi surfaces at $E_\mathrm{F}=100$meV in the vicinity of a $K$ point in the BZ of the GNR shown in a). We contrast the situation without strain ($w=0$, red circle) to externally imposed homogeneous strain ($w=0.015$, and $0.024$, blue circles). The green lines represent the quantized transverse momentum values of the unstrained GNR.}
\end{figure}

\section{\label{sec:hamiltonian}Modeling of strained graphene nanoribbons}

\subsection{Hamiltonian}
We consider a narrow and long strained GNR, clamped to unstrained graphitic leads and suspended over metallic contacts. The ribbon is assumed to have free-standing armchair edges along the transport direction $y$, and contacts with bulk electrodes along the $x$-axis, as sketched in Fig.~\ref{fig1}(a).
Such ribbons can be obtained by the oriented growth on patterned SiC substrates, \cite{SiC} etching of graphene samples with catalytic nanoparticles, \cite{etching} or by using chemical derivation. \cite{chemical}
Within the tight-binding model, the ribbon can be described by the Hamiltonian~\cite{castro-neto1}
\begin{align}
\mathcal{H}=\sum_i V_i c^{\dagger}_i c_i+\sum_{\left< ij\right>} \gamma_{ij}c^{\dagger}_i c_j,
\label{eqn-hamiltonian}
\end{align}
where $c_i$ is a fermionic annihilation operator acting on a site $i$ while $\left< ij \right>$ denotes pairs of nearest neighbors. In pristine, unstrained graphene with carbon-carbon bond lengths $r=1.42\AA$, the hopping matrix elements take the constant value $\gamma_{ij}=\gamma_0\approx-3$eV.
The system can be doped via electrostatic gates, which induce a potential step of size $V$ at the contacts. We account for this effect in Eq.~(\ref{eqn-hamiltonian}) by setting the on-site potential in the leads to $V_{i}=V$, while $V_i=0$ in the central region of an unstrained system.
For strained monolayer membranes, both the on-site potential $V_i$ as well as the hopping matrix elements $\gamma_{ij}$ are modified by the deformation of the lattice. The on-site potential then acquires an additional contribution
\begin{equation}
V_i=\frac{1}{2}r\frac{\partial \epsilon_{c}}{\partial r}\mathrm{div} \boldsymbol{u}(\boldsymbol{r}_{i}),
\end{equation}
where $\boldsymbol{u}=(u_x,u_y)$ is the displacement field of the membrane and $\epsilon_{c}$ a characteristic energy function. This contribution vanishes for homogeneous strain, and furthermore is typically well screened by the electrons in the flake and in the electrostatic environment.  \cite{marcin} We therefore focus on the  hopping matrix elements, which
must now be renormalized to \cite{castro_neto}
\begin{equation}
\gamma_{ij}= \gamma_0 e^{\eta_0 (l_{ij}/r-1)}, \quad l_{ij}\simeq r(1+ \boldsymbol{n}_{ij} \cdot \boldsymbol{\hat{w}}\boldsymbol{n}_{ij}).
\label{eqn-hopping_renormalization}
\end{equation}
Here $l_{ij}$ is the
strain-modified distance between lattice sites,
$\eta_0=\frac{\partial \gamma_0}{\partial r} \frac{r}{\gamma_0}\approx -3$ relates the change of the nearest neighbor coupling to the change of the bond length, \cite{john} $\boldsymbol{\hat{w}}$ is the $2\times 2$ strain tensor $w_{\alpha\beta}=\frac{1}{2}(\partial_{\alpha} u_{\beta}+\partial_{\beta} u_{\alpha})$ with $\alpha,\beta=x$ or $y$, and $\boldsymbol{n}_{ij}=(0,1)$, $(\frac{\sqrt{3}}{2},-\frac{1}{2})$, $(-\frac{\sqrt{3}}{2},-\frac{1}{2})$ are the unit vectors along the carbon-carbon bonds in the unstrained honeycomb lattice.

The strain-induced asymmetry in the hoppings between neighboring carbon sites is equivalent to the effect of a valley-dependent gauge vector potential  \cite{guinea}
\begin{align}
e\boldsymbol{\mathcal{A}}=\xi\frac{\hbar \eta_0}{2 r} \left(
\begin{array}{c}
w_{xx} -w_{yy}  \\
-2w_{xy} \end{array} \right), \label{eqn-vector_potential}
\end{align}
written for the states near one of the corners of the BZ, where $\xi=1$ ($\xi=-1$) for valleys $K$ ($K'$).

\subsection{Homogeneous strain}
For an externally imposed homogeneous deformation, where the GNR is elongated along the $y$-axis, the elements of the strain tensor are $w_{xx}=-\sigma w$, $w_{yy}=w$, and $w_{xy}=0$, where $\sigma=0.165$ is the Poisson ratio for graphite~\cite{proctor} and $w$ parameterizes tensile strain. In this case, both the scalar and vector potentials $V_{i}$ and $\boldsymbol{\mathcal{A}}$ are constant. The scalar potential  merely introduces a shift of the energy scale, which cannot be distinguished from the effect of electrostatic gating. The vector potential shifts the nonequivalent Dirac cones from the $K$ and $K'$ corners of the BZ into opposite directions, \cite{guinea} as shown in Fig.~\ref{fig1}(b). Infinitely wide samples are robust against such deformations and their spectrum remains gapless for strains below $20\%$. \cite{castro_neto} In contrast, GNRs behave markedly different due to quantum confinement effects, which allow for an opening of the gap even for small strains~($w \ll 20\%$). \cite{yang,lu,sena}

Figure~\ref{fig1}(c) shows a comparison between the Fermi surfaces around a $K$ point in the BZ, for an unstrained ribbon ($w=0$, red circles) and homogeneously strained ribbons ($w=0.015$ and $0.024$, blue circles)  of width $W\simeq40$nm, at energy $E_\mathrm{F}=100$meV from the DP. When the strain is smoothly increased from $w=0$ to $0.024$, the DP (black dot) crosses several quantized momenta lines (green lines) and the system undergoes multiple semiconducting-metallic-semiconducting phase transitions. Therefore, the size of the gap in the spectrum of armchair GNRs is controllable by the amount of deformation, \cite{yang,lu} within a range determined by the width of the ribbon.

\subsection{Inhomogeneous strain} To model a more realistic deformation, we assume that a suspended ribbon is clamped at the leads and stretched along the $y$-axis. Because of the clamping, the resulting deformation is inhomogeneous. \cite{manes} We neglect spontaneous wrinkling of the ribbon \cite{cerda, f_note} and consider this simplified problem within two-dimensional linear elasticity theory. \cite{timoshenko} With the origin of the coordinate system chosen in the center of the ribbon, the displacement is then prescribed by two equations, \cite{marcin}
\begin{align}\begin{split}\label{eqn:elasticity}
2\partial_{xx}u_{x}+(1-\sigma)\partial_{yy}u_{x}+(1+\sigma)\partial_{xy}u_{y}=0, \\
2\partial_{yy}u_{y}+(1-\sigma)\partial_{xx}u_{y}+(1+\sigma)\partial_{xy}u_{x}=0,
\end{split}\end{align}
accompanied by  clamped boundary conditions for the left and right edge as well as free boundary conditions for the top and bottom edge,
\begin{align}\begin{split}\label{eqn:boundary}
\mbox{clamped}&\left\{
\begin{array}{l} u_{x}(x,\pm L/2)=0 \\ u_{y}(x,\pm L/2)=\pm\frac{1}{2}wL \end{array} \right.,
\\
\mbox{free}&\left\{ \begin{array}{l} [\partial_{x}u_{x}+\sigma\partial_{y}u_{y}]_{x=\pm\frac{W}{2}} = 0
\\{} [\partial_{x}u_{y}+\partial_{y}u_{x}]_{x=\pm\frac{W}{2}} = 0 \end{array}
\right. .
\end{split}\end{align}

Despite its simplicity, the problem of finding the displacement field satisfying Eqs.~\eqref{eqn:elasticity} and \eqref{eqn:boundary} does not have an analytic solution, so that we apply the finite element method \cite{zienkiewicz_2000} with a nine-point element to determine $\boldsymbol{u}(x,y)$. Having obtained the displacement, \cite{f_note} we calculate numerically the vector potential $\boldsymbol{\mathcal{A}}(x,y)$ as predicted by the continuum model Eq.~\eqref{eqn-vector_potential}. The corresponding pseudo-magnetic field $\boldsymbol{\mathcal{B}}(x,y)=\mathrm{rot}\boldsymbol{\mathcal{A}}(x,y)$ in the $K$ valley of a GNR with width $W\simeq 40$nm, aspect ratio $L/W=4$ and inhomogeneous tensile strain $w=0.05$  in the central part
is illustrated in Fig.~\ref{fig1}(a). The pseudo-magnetic field is the largest positive~(blue) or negative~(red) near the contacts at the right and left ends, and is small in the middle part of the ribbon, where the strain is approximately homogeneous. Such strong pseudo-magnetic fields can lead to the quantization of electronic states into LLs and the appearance of gaps in the electronic spectrum.\cite{guinea,marcin,crommie} Furthermore, these field should be capable to deflect the electrons into states that are inaccessible at homogeneous strain.
In the following section, we explore these effects via the transport properties of the GNR.

\section{\label{sec:conductance}Conductance}

Having established the effects of both homogeneous and inhomogeneous strains on the electronic structure of the GNRs, we now turn to the main goal of this paper and discuss the conductance of the two-terminal device sketched in Fig.~\ref{fig1}.
In our numerical procedure, we first map the displacement directly onto the crystalline lattice of the ribbon and calculate the positions of the carbon atoms after the deformation. We then recalculate the nearest-neighbor couplings according to Eq.~\eqref{eqn-hopping_renormalization} and use this information as input for the tight-binding hamiltonian \eqref{eqn-hamiltonian}. As mentioned above, we ignore the on-site scalar potential $V_{i}$ as it is screened by the electrons in the flake and the electrostatic environment. \cite{marcin}

The phase-coherent transport properties of such two-terminal devices are encoded in the scattering matrix \cite{beenakker,henning2,blanter}
\begin{align}
S=\left( \begin{array}{cc}
                 r&t' \\
                 t&r' \end{array}\right), \label{eqn-scatering_matrix}
\end{align}
which we evaluate using the recursive Green's function technique \cite{datta,henning1} applied to the tight-binding model. Here, $t$, $t'$~($r$, $r'$) are the transmission~(reflection) amplitudes of charge carriers incident from the source or the drain leads respectively. Using the Landauer-B{\"u}ttiker formalism, \cite{buttiker} we calculate the  conductance at zero temperature
\begin{align}
G\left(E_\mathrm{F},T=0\right)&= \frac{2 e^2}{h} \mathrm{Tr}(t^{\dagger} t),
\label{eqn-cond}
\end{align}
as a function of the Fermi level $E_\mathrm{F}$. We also consider the effects of finite temperatures,  where
\begin{align}
G\left(\mu, T\right)&= \frac{2 e^2}{h} \int d E_\mathrm{F} \left( -\frac{\partial f(E_\mathrm{F}-\mu)}{\partial E_\mathrm{F}}\right) \mathrm{Tr}(t^{\dagger} t).
\label{eqn-cond1}
\end{align}
Here $\mu$ is the chemical potential, which enters together with the temperature into the Fermi distribution $f(\varepsilon)=(1+\exp(\varepsilon/k_\mathrm{B} T))^{-1}$.

Throughout the following, we set the height of the gate-controlled potential-energy step between the doped graphene leads and the suspended part
to $V=-200$meV. The resulting device is a $p$-$p'$-$p$ junction ($E_\mathrm{F}<-200$meV), an $n$-$p$-$n$ junction ($-200<E_\mathrm{F}<0$meV), or  an $n$-$n'$-$n$ junction ($E_\mathrm{F}>0$meV). In such systems, most of the conductance features are determined by scattering from the strain-modified $p$-$p'$, $n$-$p$, or $n$-$n'$ interfaces, a behavior which can be investigated by analyzing the spatial distribution of the electronic states. Within the used formalism, this can be revealed via the local density of states, \cite{gasparian}
\begin{align}
\mathrm{LDOS}= \frac{i}{4 \pi} \mathrm{Tr}\left( S^{\dagger} \frac{\partial S}{\partial V_i}-\frac{\partial S^{\dagger}}{\partial V_i} S\right),
\label{eqn-LDOS}
\end{align}
which corresponds to the response of the scattering amplitudes to a small local perturbation $\delta V_i$ added to the Hamiltonian in Eq.~(\ref{eqn-hamiltonian}).

\subsection{\label{subsec:homogeneous}Transport across homogeneously strained armchair GNRs}

\begin{figure*}[t!]
\includegraphics[width=2\columnwidth]{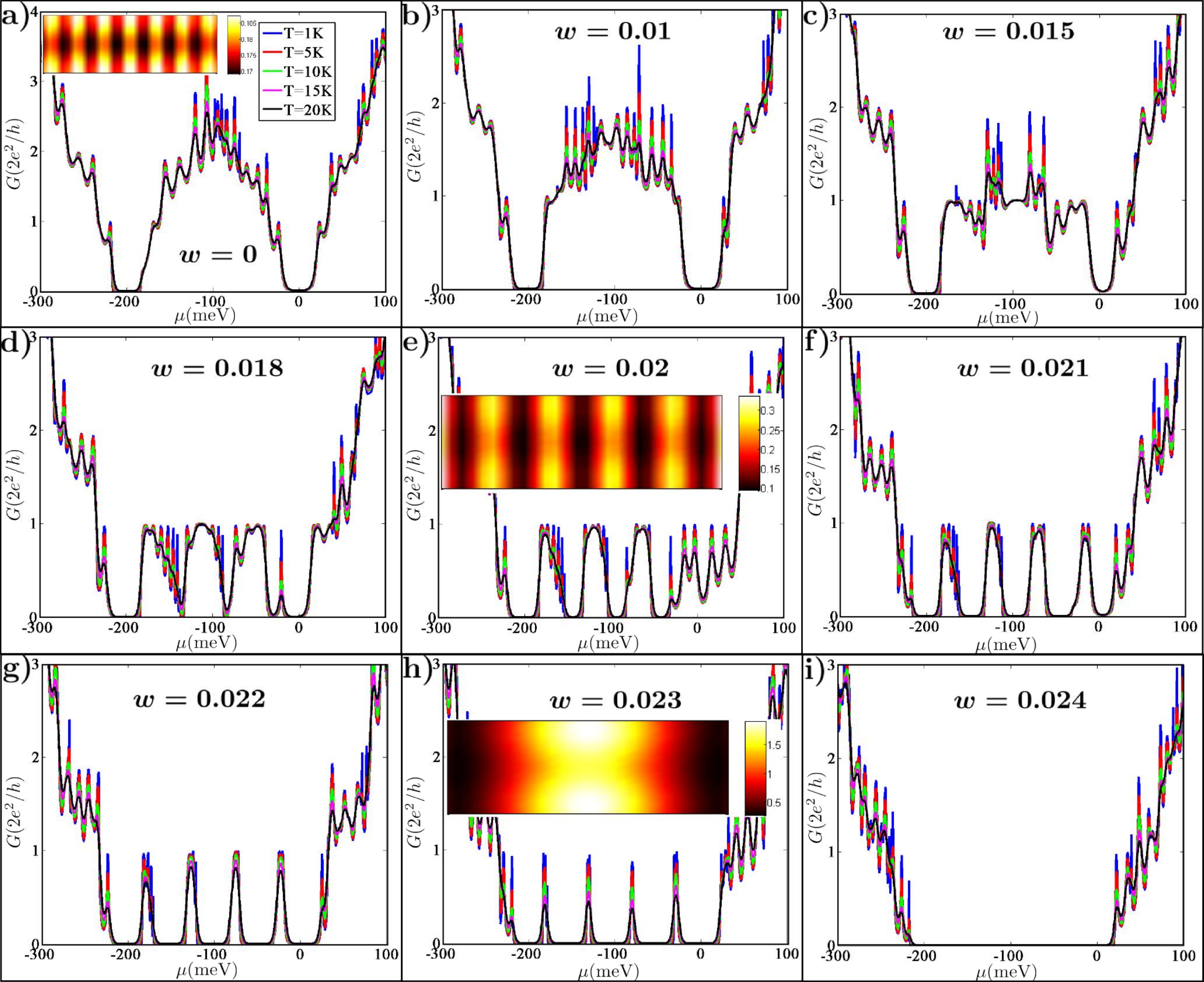}
\caption{\label{fig2}(Color online) Linear-response conductance as a function of chemical potential $\mu$ at several fixed temperatures $T$, for a homogeneously strained GNR of width $W\simeq40$nm and aspect ratio $L/W=3$. The leads are doped via an onsite potential $V=-200$meV.
Each panel corresponds to a different value of externally imposed homogeneous strain. Panels a), e) and h) also show the spatial structure of electron wave amplitudes at energy $E_\mathrm{F}=-129.2$meV, evaluated using Eq.~(\ref{eqn-LDOS}). }
\end{figure*}

Figure \ref{fig2} shows the numerically evaluated conductance Eq.~(\ref{eqn-cond1})
for a GNR of width $W\simeq40$nm and aspect ratio $L/W=3$, as a function of chemical potential, for various values of homogeneous strain  and temperature.

The unstrained GNR~[Fig.~\ref{fig2}(a)] is semiconducting with a gap of $\simeq 30$meV, as determined by the  quantization of the transverse momentum discussed above. The conductance exhibits two minima at $\mu=-200$meV and $0$meV and a local maximum at $-100$meV. The conductance oscillations away from the two DPs are due to the Fabry-P\'{e}rot-like standing-wave resonances in the electron transmission across the potential barrier. \cite{tudorovsky,katsnelson,pereira} This can be seen in the LDOS profile shown in the insert, which we calculated using Eq.~(\ref{eqn-LDOS}) at the resonance energy $E_\mathrm{F}=-129.2$meV.

For the homogeneously strained GNRs in Figs.~\ref{fig2}(b)-(i), the results show that the conductance continues to exhibit the minima at $\mu=-200$meV and $0$meV. The effect of the strain is most noticeable in the energy range $-200$meV$<\mu<0$meV, where the system constitutes an $n$-$p$-$n$ junction.

For deformations $w<0.015$~[Figs.~\ref{fig2}(b)-(c)], the conductance in this range is broadly suppressed. This can be attributed to the  strain-induced shift of the DP which results in a misalignment between the Fermi surfaces in the unstrained leads and the strained central region, as illustrated by the example in Fig.~\ref{fig1}(c). Only the quantized momenta that cross the overlapping area of the two Fermi surfaces correspond to propagating modes in the leads that couple to propagating modes in the suspended region and therefore contribute towards transport. With increasing strain, the area of the overlap decreases, and the conductance is reduced as an increasing number of conducting channels become blocked.

For strain $0.018 \leq w <0.024$~[Figs.~\ref{fig2}(d)-(h)] the conductance exhibits a series of well defined resonances. In this range of strain, the area of the overlap between the Fermi surfaces in the leads and in the central region is narrower than the separation between neighboring quantized momenta lines. For a fixed strain $w$, the width of the overlap remains constant with varying energy, but the overlap itself is shifted in the momentum plane along the $k_x$-axis. Therefore,  zero-conductance plateaus appear periodically in the range of energies when there is no quantized-momentum line crossing the area of the overlap. In this case, the propagating modes in the central device only couple to evanescent modes in the leads, leading to the formation of transport gaps in the system. The finite-conductance resonances are entirely due to Fabry-P\'{e}rot-like standing wave patterns, as illustrated by the LDOS profiles in Fig.~\ref{fig2}(e, h).

For strains $w \geq 0.024$~[Fig.~\ref{fig2}(i)], the conductance in the range $-200$meV$<\mu<0$meV is completely suppressed, which results from the
complete misalignment between the Fermi surfaces in the two regions at such  strong deformations. \cite{pereira} This threshold for the insulating behavior is controlled by the parameters used in Fig.~\ref{fig2} and can be lowered (raised) by reducing (increasing) the height of the potential step $V$ between the central part of the ribbon and the contacts.

The finite-conductance resonances are characteristic for junctions between regions of different polarity~($n$-$p$-$n$ junctions) and are absent in junctions between regions of the same polarity ($n$-$n'$-$n$ and $p$-$p'$-$p$ junctions). This is because for $\mu<-200$meV and $\mu>0$meV the region of overlap  of the Fermi surfaces increases with increasing energy, and contains an increasing number of quantized momentum lines. With larger strains ($w > 0.03$) the two Fermi surfaces will only start overlapping at energies further away from the DPs ($E_\mathrm{F}<-200$meV or $E_\mathrm{F}>0$meV), which results in a widening of the transport gap in Fig.~\ref{fig2}(i). For example, at $w=0.05$ we find that the  conductance $G$  vanishes in the entire energy range $|E_\mathrm{F}|\leq100$meV around the DP of the suspended region.

\begin{figure}[t!]
\includegraphics[width=\columnwidth]{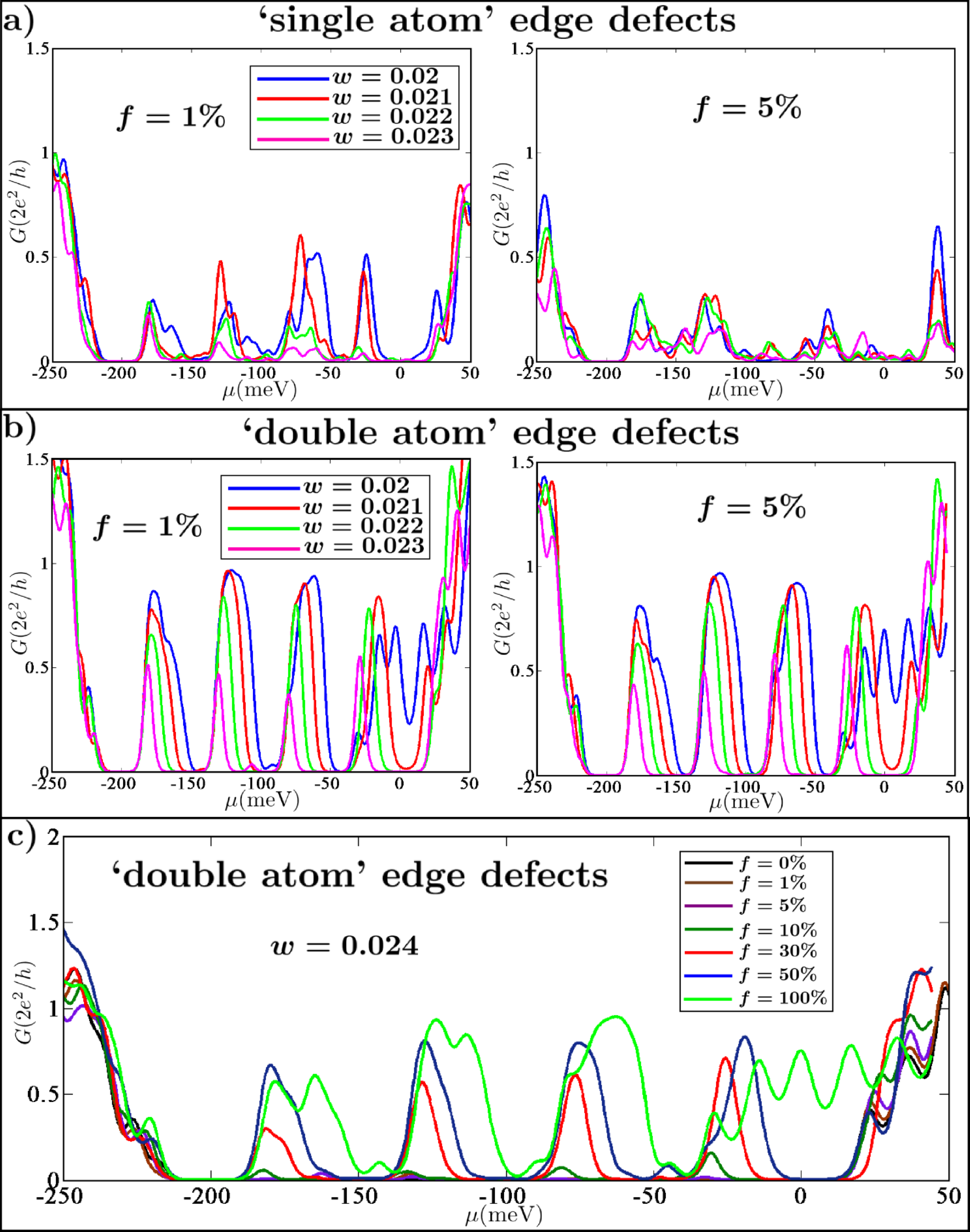}
\caption{\label{fig3}(Color online) Linear-response conductance $G$ as a function of chemical potential $\mu$ at a fixed temperature $T=20$K, for homogeneously strained GNRs, of width $W=40$nm and aspect ratio $L/W=3$ subjected to various types of disorder. Panel (a) shows the effect of $f=1\%$ and $5\%$  single-atom vacancies for strain $w=0.02$, $0.021$, $0.022$, and $0.023$, while panel (b) shows the corresponding effect of double-atom vacancies. Panel (c) shows the conductance for various fractions of  double-atom vacancies for a fixed strain $w=0.024$.}
\end{figure}

\subsection{\label{subsec:disorder}Influence of edge disorder in GNRs}
Ideal ribbons with perfectly cut edges are not realistic, as most fabricated structures present a certain degree of roughness at the edges. \cite{xli,kim} Therefore, in this subsection we establish the robustness of the strain-induced conductance resonances against edge defects. We introduce edge disorder by randomly removing a fraction $f$ of individual atoms within a strip of width $2r$ from the edges  in the strained region (single-atom vacancies), \cite{henning1,li,evaldsson,castro-neto2} and compare this to the removal of carbon-carbon dimers in the outer-most rows of the edges (double-atom vacancies). \cite{li,areshkin} The missing atoms are modeled by setting all the nearest neighbor hopping elements $\gamma_{ij}$ to zero.

Figure~\ref{fig3} shows the effect on the conductance at a fixed temperature $T=20$K for the homogeneously strained GNR of width $W\simeq40$nm and aspect ratio $L/W=3$, for several strains $w$ and percentages $f$ of single-atom and double-atom vacancies as indicated in each panel.
Panel (a) shows that single-atom defects induce a smearing and suppression of the finite-conductance resonances.  Previous studies have shown that in the absence of strain, such edge disorder gives rise to drastic changes in the transport properties of armchair GNRs, by inducing large fluctuations in the conductance even for small percentages of defects. By breaking the sublattice symmetry \cite{li} and acting as short-range scatterers, \cite{evaldsson,castro-neto2} such edge defects induce backscattering, Anderson-type localization, and even the formation of conduction gaps. Similarly, our calculations show that the conductance rapidly degrades with increasing edge disorder, as an increasing number of conductive paths become blocked.
As compared to the results for a defect-free system, Fig.~\ref{fig2}(d)-(h), the conductance is already greatly reduced in the presence of $f=1\%$ edge disorder, and the resonances become barely visible when $f=5\%$.

\begin{figure*}[t!]
\includegraphics[width=2\columnwidth]{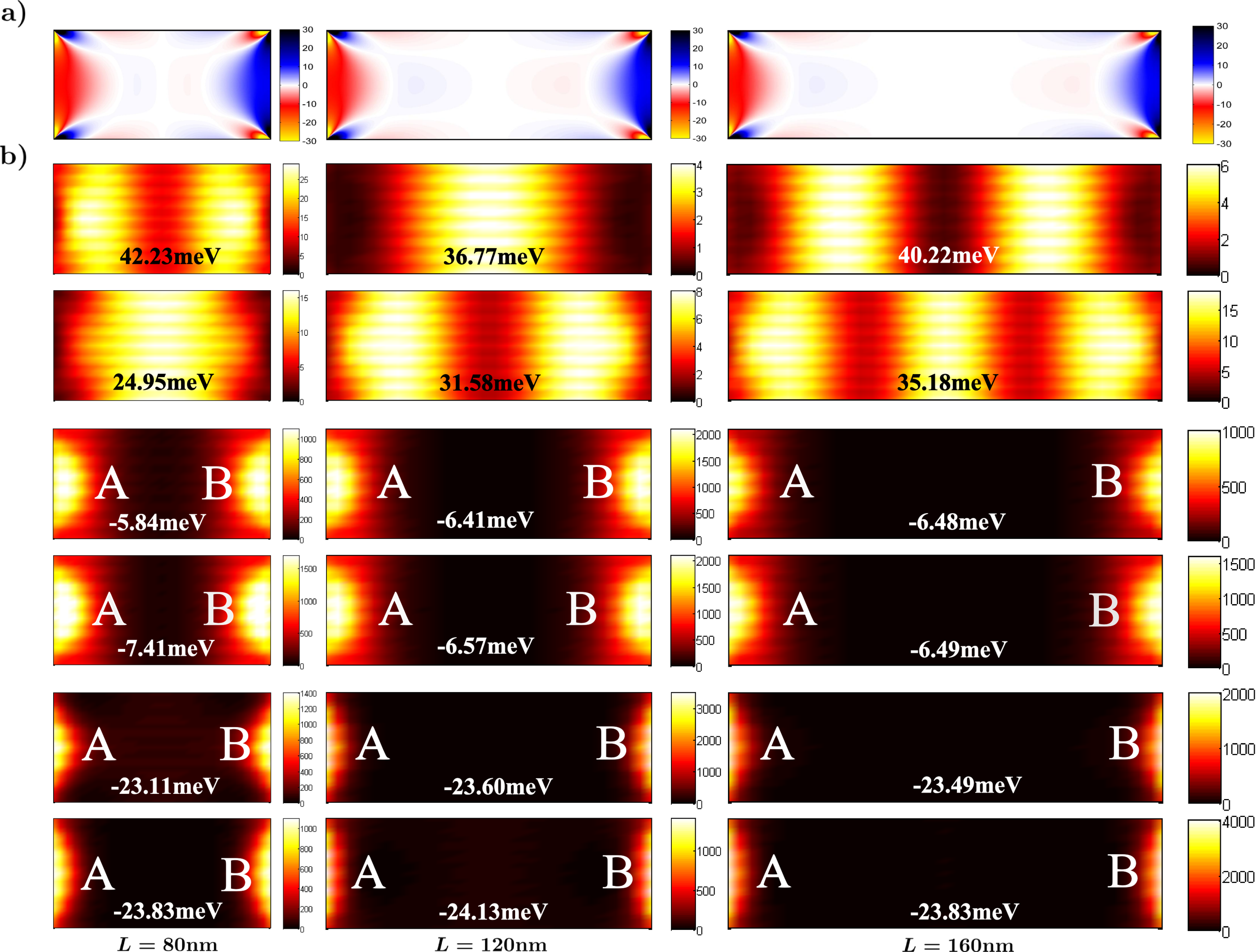}
\caption{\label{fig4}(Color online) (a) Suspended GNRs of width $W\simeq40$nm and aspect ratios $L/W=2$, $3$, and $4$, which are clamped at the highly-doped contacts. The color code shows the strength of the pseudo-magnetic fields $\mathcal{B}(\mathrm{Tesla})$ for electrons in the $K$ valley, for
inhomogeneous strain $w=0.05$.   (b) Spatial structure of resonance states for selected resonances identified in Fig.~\ref{fig5}.}
\end{figure*}

Double-atom edge defects, on the other hand, preserve the sublattice symmetry and therefore are expected to induce only  small changes in the conductance. \cite{li} This is confirmed by the results in Fig.~\ref{fig3}(b). Compared to the defect-free ribbon, Fig.~\ref{fig2}(d)-(h), the conductance for $f=1\%$ and $f=5\%$ disorder shows remarkably little changes. Even at higher degrees of disorder, the resonances are still visible. The most significant effect is obtained for $w=0.024$ strain, depicted in Fig.~\ref{fig3}(c), where we show the conductance calculated for various percentages of edge disorder. In this case, the transport properties of the device observed in the ballistic regime are restored by large percentages of double-atom edge defects. This behavior can be understood by comparing the two theoretical disorder extremes: $f=0\%$ and $f=100\%$. At $f=0\%$~(no edge disorder), the central device and leads are perfectly matched, both having a width $W$ and the same transverse momentum quantization. The requirement for conservation of transverse momenta leads to the complete suppression of the conductance since the Fermi surfaces in the leads and the strained suspended region do not overlap. At $f=100\%$ edge disorder the outermost rows of dimer lines at the top and bottom edges of the suspended region are completely removed. Therefore this region has a smaller width and correspondingly different quantized transverse momenta than the leads. This mismatch induces a mode mixing mechanism at the interfaces with the contacts, leading to the appearance of finite-conductance resonances even if the Fermi surfaces do not overlap. Other degrees of edge disorder will induce a random mixture of local boundary conditions at the edges,  \cite{areshkin} and therefore yield intermediate conductance results.

\begin{figure*}[t!]
\includegraphics[width=2\columnwidth]{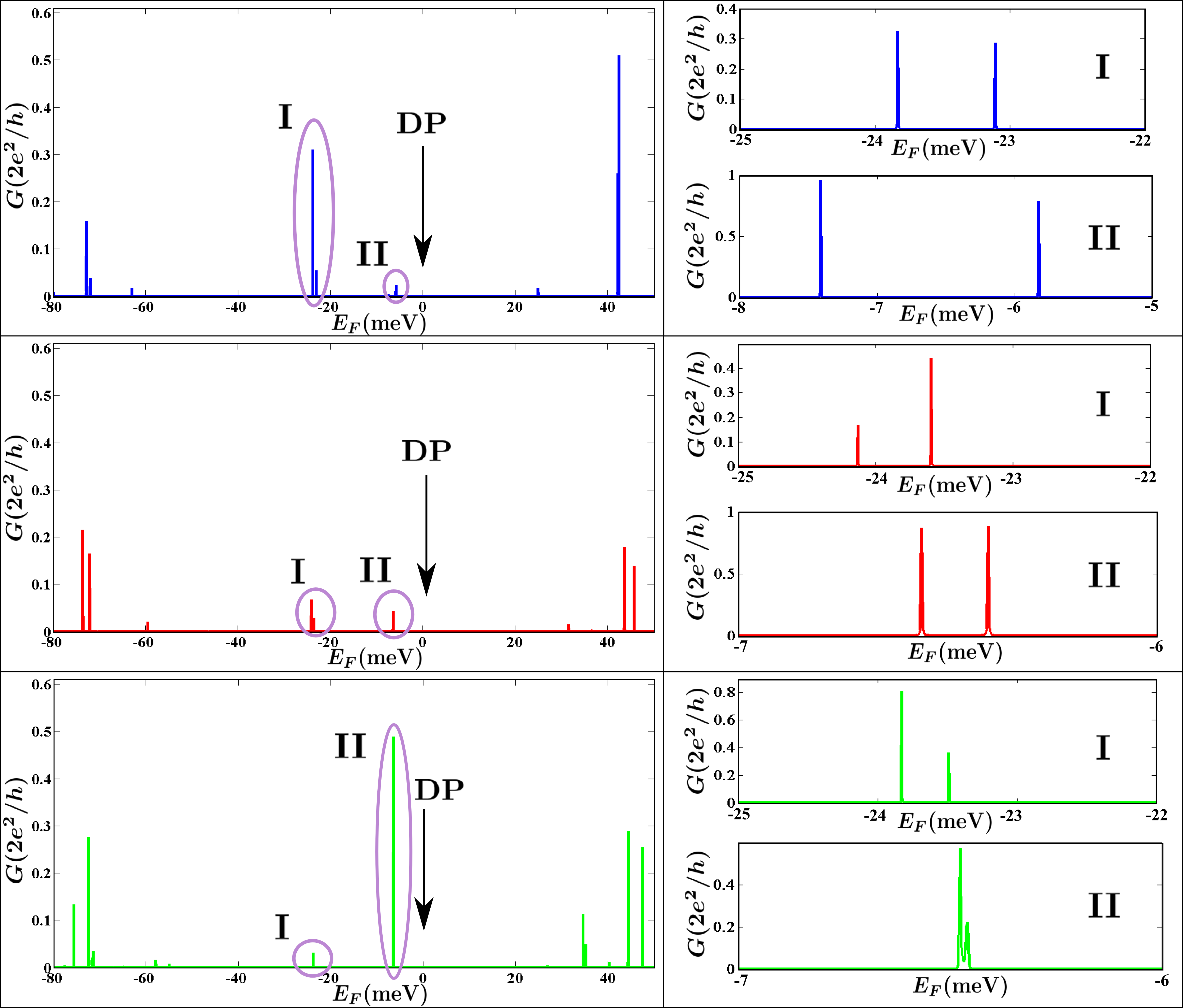}
\caption{\label{fig5}(Color online) Left: Zero-temperature conductance $G$ as a function of Fermi energy for the GNRs shown in Fig.~\ref{fig4} (top: $L/W=2$; middle: $L/W=3$; bottom: $L/W=4$). Right: Highly resolved results for the groups of peaks identified in the panels on the left.}
\end{figure*}

\subsection{\label{subsec:inhomogeneous}Transport across inhomogeneously strained armchair GNRs}
We now study the transport in suspended graphene nanoribbons, which display inhomogeneous strain distributions. In contrast to clean and disordered homogeneously strained GNRs, where the conductance vanishes around the neutrality point of the suspended part, we now find that the conductance features several additional resonances, including resonances close to the neutrality point. \cite{gradinar} Since previous works  predict the formation of pseudo magnetic LLs in such systems, \cite{marcin,guinea,crommie} we aim to determine whether any of the observed new features in the conductance reflect this quantization of the electronic states. We focus our study on the energy range $|E_\mathrm{F}|<100$meV around the DP in the suspended region, where, if present, the first few LLs are well resolved. Outside of this energy range, the states are likely to be broadened and smeared. \cite{guinea}

We consider three inhomogeneously strained ribbons, of width $W\simeq40$nm and aspect ratios $L/W=2$, $3$ and $4$. The  pseudo-magnetic field distributions for inhomogeneous tensile strain $w=0.05$ are shown in Fig.~\ref{fig4}(a). Using Eq.~(\ref{eqn-cond}), we calculate the zero-temperature conductance, which is shown in the left panels of  Fig.~\ref{fig5}. In contrast to the results obtained in the previous subsections, where the conductance was completely suppressed for homogeneous strains $w\geq0.024$, here we find four groups of sharp and clearly defined resonance conductance peaks for each considered aspect ratio. The two groups positioned furthest from the DP, at $E_\mathrm{F}\simeq-70$meV and $\simeq40$meV, contain several resonances with their number being proportional to the aspect ratio of the respective ribbons. For the other two groups, positioned in the energy range $-25$meV$<E_\mathrm{F}<0$meV just below the DP, the highly resolved conductance results in the right panels in Fig.~\ref{fig5} reveal that these resonances always occur in pairs of two. Furthermore, the splitting of the two peaks in each group decreases with increasing aspect ratio.

To uncover the origin of each group of peaks, we analyze the spatial distribution of the corresponding electronic states using Eq.~(\ref{eqn-LDOS}), and arrive at the LDOS profiles shown in Fig.~\ref{fig4}(b).
As illustrated in the top two rows, the states away from the DP correspond to Fabry-P\'{e}rot-like standing waves that form due to multiple electron reflections from the left and right interfaces. Similarly to the LDOS profiles in Fig.~\ref{fig2}, such states are confined to the central part of the structure, where the strain distribution is approximately homogeneous. The inhomogeneity near the contacts is still important as it mixes states with different transverse momentum and thus allows  the charge carriers to overcome the misalignment of the Fermi surfaces described in Sec.~\ref{sec:hamiltonian}. For the two groups in the energy range $-25$meV$<E_\mathrm{F}<0$meV, where the resonances occur in almost degenerate pairs, the LDOS profiles shown in the bottom four rows of Fig.~\ref{fig4}(b) do, however, point towards a very different behavior. Unlike any of the resonances we found up to now, the spatial structure for these states clearly resembles the pseudo-magnetic field distributions, which is an indicator for the formation of LLs.

As demonstrated next, this quadruplet of resonances (two groups, each containing two conductance peaks) can be attributed to the $n=0$ pseudo-magnetic Landau level induced by the inhomogeneity at the interfaces.
We exploit a unique feature of this LL in armchair GNRs, namely that the electron amplitude resides either on the $A$ or $B$ sublattice.
\cite{guinea,suzura,prada,manes2,ando,schomerus,poli}
This sublattice polarization can be seen from the low-energy Hamiltonian  \cite{castro-neto1}
\begin{align}
\begin{array}{cc}
 H=v_\mathrm{F}\left(\begin{array}{cc}
    0 & \boldsymbol{\hat{\pi}}^{\dagger}  \\
    \boldsymbol{\hat{\pi}} & 0 \end{array}\right),& \begin{array}{c}\boldsymbol{\hat{\pi}}=\boldsymbol{\hat{p}}+\frac{e}{c}\boldsymbol{\mathcal{A}}, \\
                                                    \boldsymbol{\hat{p}}= p_x+i p_y, \end{array}
 \end{array}.
\end{align}
Here $v_\mathrm{F}$ is the Fermi velocity, $\boldsymbol{\hat{p}}$ parameterizes the in-plane momentum relative to the $K$ or $K'$ point and $\boldsymbol{\mathcal{A}}$ is the vector potential in Eq.~(\ref{eqn-vector_potential}). The operator $\boldsymbol{\hat{\pi}}$ fulfills $[\boldsymbol{\hat{\pi}}, \boldsymbol{\hat{\pi}}^\dagger]=\mbox{const}$ and acts as an annihilation operator if the pseudomagnetic field is positive.
Acting by this Hamiltonian on the state
\begin{align}
\left(\begin{array}{c}
    \left|0\right\rangle  \\
    0 \end{array}\right),
\end{align}
where $\boldsymbol{\hat{\pi}}\left|0\right\rangle=0$,
we obtain the eigenvalue $E=0$. This eigenstate has a finite amplitude on the $A$ sublattice, but vanishing amplitude on the $B$ sublattice.
For negative value of the pseudomagnetic field the sublattice polarization moves onto the B sublattice.
However, in all cases the selected sublattice is independent of the valley. \cite{guinea} In contrast, higher order LLs and Fabry-P\'{e}rot-like resonances occupy both sublattices equally. \cite{ando}

\begin{figure*}[t!]
\includegraphics[width=2\columnwidth]{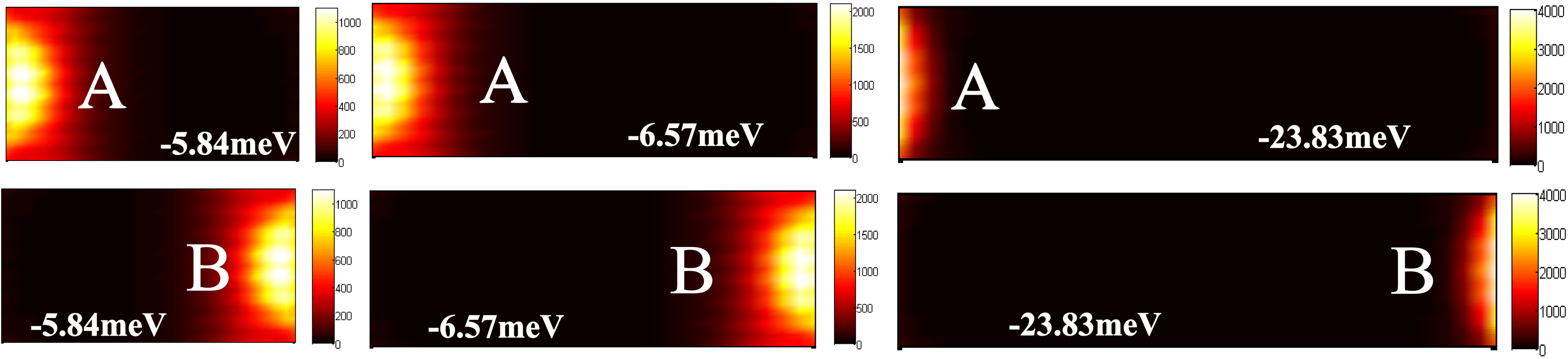}
\caption{\label{fig6} Sublattice-resolved electron amplitude for the resonances in Fig.~\ref{fig5}~($L/W=2$, $3$, and $4$, at energies $E_{\mathrm{F}}=-5.84$meV, $-6.57$meV, and $-23.83$meV respectively), obtained from Eq.~(\ref{eqn-LDOS}) by placing the probing perturbation $\delta V_i$ onto $A$ sites (top panels) or onto $B$ sites (bottom panels).}
\end{figure*}

By placing the probing perturbation $\delta V_i$ in Eq.~(\ref{eqn-LDOS}) on either the $A$ or on the $B$ sites, we find that the low-energy resonances are localized on the $A$-sites near the left interface (where $\mathcal{B}<0$), and on the $B$-sites near the right interface (where $\mathcal{B}>0$).
This is illustrated in Fig.~\ref{fig6}.

Further evidence that supports our interpretation of the origin of these states is the fact that we find four such low-energy resonances, with the separation between each pair inversely proportional to the aspect ratio of the ribbon. The $x\rightarrow -x$ reflection symmetry of the system maps the $K$ and $K'$ valleys onto each other, which results in the formation of a symmetric and an anti-symmetric superposition of the two valley manifestations of the $n=0$ LL. This leads to a splitting of the $n=0$ LL into two branches, corresponding to each of the two groups of resonances. The branch located at $E_\mathrm{F} \approx-24$meV is valley-symmetric and displays a maximum on the symmetry axis. The  branch located at $E_\mathrm{F} \approx-7$meV is valley-antisymmetric and displays a nodal line on the symmetry axis. Two states appear in each of the branches due to the hybridization of states localized at the two contacts.
The tunnel coupling of these states is provided by the evanescent tails of the electronic wave functions in the central part of the system
where $\mathcal{B}$ is small. Naturally, since the overlap of the evanescent tails decreases with increasing aspect ratio, the splitting in each of these pairs is smaller in a longer ribbon, thus explaining the trend we highlighted in our discussion of the highly resolved conductance result in the right panels of Fig.~\ref{fig5}.


\section{\label{sec:conclusions}Conclusion}
In conclusion, we performed a systematic study of the transport characteristics of homogeneously and inhomogeneously strained suspended armchair graphene nanoribbons. The combination of strain-induced shifts of the Dirac point in the momentum plane and size-confinement effects leads to significant modifications in the transport of homogeneously strained systems. In particular, an uncommon resonance structure appears when both of these effects compete. Large percentages of single-atom vacancies destroy the observed resonant structure. In contrast, `double-site' vacancies do not suppress the conductance, and even can restore the ballistic transport properties. For inhomogeneous deformations, we have found that the inhomogeneity developed near the contacts aids the resonant transmission of charge carriers, either through a mode mixing mechanism or through tunneling via the sublattice-polarized $n=0$ pseudo-magnetic Landau level. The mode mixing leads to the coupling to Fabry-P\'{e}rot-like standing waves in the central part of the ribbon, which results in the formation of additional conductance peaks far from the Dirac point. The states associated to the $n=0$ pseudo-magnetic Landau level form near the contact regions, and give rise to two pairs of conductance peaks near the Dirac point.

\section{Acknowledgements}
We thank F.~Guinea and H.~Ochoa for useful discussions. This project was funded by European Graphene Flagship Project, EPSRC S\&IA grant, ERC Synergy Grant "Hetero2D", and by the Royal Society Wolfson Research Merit Award.


\begin{thebibliography}{99}
\bibliographystyle{apsrev}
\bibitem{castro-neto1} A.~H.~Castro-Neto, F.~Guinea, N.~M.~R.~Peres, K.~S.~Novoselov, and A.~K.~Geim, Rev. Mod. Phys. \textbf{81}, 109 (2009).
\bibitem{castro_neto} V. M. Pereira, A. H. Castro Neto, and N.~M.~R.~Peres, Phys. Rev. B \textbf{80}, 045401 (2009).
\bibitem{ribeiro}
R.~M.~Ribeiro, V.~M.~Pereira, N.~M.~R.~Peres, P.~R.~Briddon, and A.~H.~Castro Neto, New J. Phys. \textbf{11}, 115002 (2009).
\bibitem{liu_ming} F.~Liu, P.~Ming, and J.~Li, Phys. Rev. B \textbf{76}, 064120 (2007).
\bibitem{kim1} K.~S.~Kim, Y.~Zhao, H.~Jang, S.~Y.~Lee, J.~M.~Kim, K.~S.~Kim, J.-H.~Ahn, P.~Kim, J.~-Y.~Choi, and B.~H.~Hong, Nature \textbf{457}, 706 (2009).
\bibitem{strong} C.~Lee, X.~Wei, J.~W.~Kysar, and J.~Hone, Science \textbf{321}, 385 (2008).
\bibitem{membrane1} T.~J.~Booth, P.~Blake, R.~R.~Nair, D.~Jiang, E.~W.~Hill, U.~Bangert, A.~Bleloch, M.~Gass, K.~S.~Novoselov, M.~I.~Katsnelson, and A.~K.~Geim, Nano Lett. \textbf{8}, 2442 (2008).
\bibitem{membrane2}
     J.~S.~Bunch, S.~S.~Verbridge, J.~S.~Alden, A.~M.~van der Zande, J.~M.~Parpia, H.~G.~Craighead, and P.~L.~McEuen,  Nano Lett.  \textbf{8}, 2458 (2008).
\bibitem{effects} Z.~H.~Ni, T.~Yu, Y.~H.~Lu, Y.~Y.~Wang, Y.~P.~Feng, and Z.~X.~Shen, ACS Nano \textbf{2}, 2301 (2008).
\bibitem{effects2} Z.~H.~Ni, H.~M.~Wang, Y.~Ma, J.~Kasim, Y.~H.~Wu, and Z.~X.~Shen, ACS Nano \textbf{2}, 1033 (2008).
\bibitem{koshelev} S.~V.~Iordanskii and A.~E.~Koshelev, Zh. Eksp. Teor. Fiz. \textbf{91}, 326 (1986).
\bibitem{guinea} F.~Guinea, M.~I.~Katsnelson, and A.~K.~Geim, Nat. Phys. \textbf{6}, 30 (2010).
\bibitem{vozmediano} M.~A.~H.~Vozmediano, M.~I.~Katsnelson, and F.~Guinea, Phys. Rep. \textbf{496}, 109 (2010)
 \bibitem{rainis}
 D.~Rainis, F.~Taddei, M.~Polini, G.~ Le\'{o}n, F.~Guinea, and V.~I.~Fal'ko, Phys. Rev. B \textbf{83}, 165403 (2011).
\bibitem{suzura} H.~Suzuura and T.~Ando, Phys. Rev. B \textbf{65}, 235412 (2002).
\bibitem{manes2}
J.~L.~Ma{\~n}es, Phys. Rev. B \textbf{76}, 045430 (2007).
\bibitem{prada}
 E.~Prada, P.~San-Jose, G.~Le{\'o}n, M.~M.~Fogler, and F.~Guinea, Phys. Rev. B \textbf{81}, 161402 (2010).
\bibitem{low} T.~Low, F.~Guinea, and M.~I.~Katsnelson, Phys. Rev. B \textbf{83}, 195436 (2011).
\bibitem{pereira} V.~M.~Pereira and A.~H.~Castro Neto, Phys. Rev. Lett. \textbf{103}, 046801 (2009).
\bibitem{crommie} N.~Levy, S.~A.~Burke, K.~L.~Meaker, M.~Panlasigui, A.~Zettl, F.~Guinea, A. H.~Castro Neto, and M.~F.~Crommie, Science \textbf{329}, 544 (2010).
\bibitem{schomerus}  H. Schomerus and N. Y. Halpern, Phys. Rev. Lett. \textbf{110}, 013903 (2013).
\bibitem{poli}  C. Poli, J. Arkinstall, and H. Schomerus, preprint arXiv:1408.3014 (2014).
\bibitem{marcin} M.~ Mucha-Kruczynski and V.~I.~Fal'ko, Solid State Comm. \textbf{152}, 1442 (2012).


\bibitem{buttiker} M.~B{\"u}ttiker, Phys. Rev. Lett. \textbf{57}, 1761 (1986).
\bibitem{henning1} J.~P.~Robinson and H.~Schomerus, Phys. Rev. B \textbf{76}, 115430 (2007).
\bibitem{datta} S.~Datta, \emph{Electronic transport in mesoscopic systems} (Cambridge University Press, Cambridge, 1988).
\bibitem{suspended} K. I. Bolotin, K. J. Sikes, Z. Jiang, M. Klima, G. Fudenberg, J. Hone, P. Kim, and H. L. Stormer, Sol. State Commun. \textbf{146}, 351 (2008).
\bibitem{suspended2}
X. Du, I. Skachko, A. Barker, and E. Y. Andrei, Nature Nanotech. \textbf{3}, 491 (2008).
\bibitem{with_strain} W. Bao, K. Myhro, Z. Zhao, Z. Chen, W. Jang, L. Jing, F. Miao, H. Zhang, C. Dames, and C. N. Lau, Nano Lett. \textbf{12}, 5470 (2012).
 \bibitem{with_strain2}
 H. Zhang, J.-W. Huang, J. Velasco Jr., K. Myhro, M. Maldonado, D. D. Tran, Z. Zhao, F. Wang, Y. Lee, G. Liu, W. Bao, and C. N. Lau, Carbon \textbf{69}, 336 (2014).
  \bibitem{with_strain3}
 C. Lee, X. Wei, J. W. Kysar, and J. Hone, Science \textbf{321}, 385 (2008).
\bibitem{gradinar} D. A. Gradinar, M. Mucha-Kruczynski, H. Schomerus, and V. I. Fal'ko, Phys. Rev. Lett. \textbf{110}, 266801 (2013).

\bibitem{SiC} J.~Hicks, A.~Tejeda, A.~Taleb-Ibrahimi, M.~S.~Nevius, F.~Wang, K.~Shepperd, J.~Palmer, F.~Bertran, P.~Le~F\`{e}vre, J.~Kunc, W.~A.~de~Heer, C.~Berger, and E.~H.~Conrad, Nat. Phys. \textbf{9}, 49 (2012).
\bibitem{etching} L.~C.~Campos, V.~R.~Manfrinato, J.~D.Sanchez-Yamagishi, J.~Kong, and P.~Jarillo-Herrero, Nano Lett. \textbf{9}, 2600 (2009).
\bibitem{chemical} X.~Li, X.~Wang, L.~Zhang, S.~Lee, and H.~Dai, Science \textbf{319}, 1229 (2008).
\bibitem{john} R.~Ferone, J.~R.Wallbank, V.~Zolyomi, E.~McCann, and V.~I.~Fal'ko, Solid State Comm. \textbf{151}, 1071 (2011).
\bibitem{proctor} O.~L.~Blakslee, D.~G.~Proctor, E.~J.~Seldin, G.~B.~Spence, and T.~Weng, J. Appl. Phys. \textbf{41}, 3373 (1970).
\bibitem{yang} Y.~Li, X.~Jiang, Z.~Liu, and Z.~Liu, Nano Res. \textbf{3}, 545 (2010).
\bibitem{lu} Y.~Lu and J.~Guo, Nano Res. \textbf{3}, 189 (2010).
\bibitem{sena} S.~H.~R.~Sena, J.~M.~Pereira Jr, G.~A.~Farias, F.~M.~Peeters, and R.~N.~Costa Filho, J. Phys.: Condens. Matter \textbf{24}, 375301 (2012).
\bibitem{manes} J. L. Ma$\mathrm{\tilde{n}}$es, F. de Juan, M. Sturla, and M. A. H. Vozmediano, Phys. Rev. B \textbf{88}, 155405 (2013).
\bibitem{f_note} Note that we have neglected the possibility of spontaneous wrinkling \cite{cerda} since strain limits their formation in suspended samples by increasing the transverse rigidity. \cite{ochoa} We have also applied a finite cutoff to regularize the formally divergent displacements~\cite{williams} one finds using linear elasticity theory~\cite{timoshenko} near the corners of the clamped ends of the ribbon.
\bibitem{cerda} E.~Cerda and L.~Mahadevan, Phys. Rev. Lett. \textbf{90}, 074302 (2003).
\bibitem{timoshenko} S.~P.~Timoshenko and J.~N.~Goodier, \emph{Theory of Elasticity} (McGraw-Hill, Singapore, 1970).
\bibitem{zienkiewicz_2000} O. C. Zienkiewicz and R. L. Taylor, \emph{The Finite Element Method}, Vol. I–III, 5th ed. (Butterworth-Heinemann, Oxford, 2000).
\bibitem{henning2} H.~Schomerus, Phys. Rev. B \textbf{76}, 045433 (2007).
\bibitem{beenakker} C.~W.~J.~Beenakker, Rev. Mod. Phys. \textbf{69}, 731 (1997).
\bibitem{blanter} Y.~M.~Blanter and M.~B{\"u}ttiker, Phys. Rep. \textbf{336}, 1 (2000).
\bibitem{gasparian} V.~Gasparian, T.~Christen, and M.~B{\"u}ttiker, Phys. Rev. A, \textbf{54}, 4022 (1996).
\bibitem{katsnelson} M.~I.~Katsnelson, K.~S.~Novoselov, and A.~K.~Geim, Nat. Phys. \textbf{2}, 620 (2006).
\bibitem{tudorovsky} T.~Tudorovskiy, K.~J.~A.~Reijnders, and M.~I.~Katsnelson, Phys. Scr. \textbf{T146}, 014010 (2012).
\bibitem{xli} X.~Li, X.~Wang, L.~Zhang, S. Lee, and H.~Dai, Science \textbf{319}, 1229 (2008).
\bibitem{kim} M.~Y.~Han, B.~{\"O}zyilmaz, Y.~Zhang, and P.~Kim, Phys. Rev, Lett. \textbf{98}, 206805 (2007).
\bibitem{li} T.~C.~Li and S.-P.~Lu, Phys. Rev. B \textbf{77}, 085408 (2008).
\bibitem{evaldsson} M.~Evaldsson, I.~V.~Zozoulenko, H.~Xu, and T.~Heinzel, Phys. Rev. B \textbf{78}, 161407 (2008).
\bibitem{castro-neto2} E.~R.~Mucciolo, A.~H.~Castro Neto, and C.~H.~Lewenkopf, Phys. Rev. B \textbf{79}, 075407 (2009).
\bibitem{areshkin} D.~A.~Areshkin, D.~Gunlycke, and C.~T.~White, Nano Lett. \textbf{7}, 204 (2007).
\bibitem{ando} Y.~Zheng and T.~Ando, Phys. Rev. B, \textbf{65}, 245420 (2002).
\bibitem{ochoa} E.~V.~Castro, H.~Ochoa, M.~I.~Katsnelson, R.~V.~Gorbachev, D.~C.~Elias, K.~S.~Novoselov, A.~K.~Geim, and F.~Guinea, Phys. Rev. Lett. \textbf{105}, 266601 (2010).
\bibitem{williams} M.~L.~Williams, J. Appl. Mech. - T. ASME \textbf{19}, 526 (1952).

\end{thebibliography}
\end{document}